\documentclass[%
 aip,
 amsmath,amssymb,
 reprint,%
]{revtex4-1}

\usepackage{graphicx}
\usepackage{dcolumn}
\usepackage{bm}
\usepackage[utf8]{inputenc}
\usepackage[T1]{fontenc}
\usepackage{mathptmx}
\usepackage{etoolbox}
\usepackage{xcolor}
\usepackage{ulem}
\DeclareRobustCommand{\erase}{\bgroup\markoverwith{\textcolor{red}{\rule[.5ex]{2pt}{0.4pt}}}\ULon}

\usepackage{here}

\newcommand{\keio}{School of Fundamental Science and Technology,
Keio University, Yokohama, Kanagawa 223-8522, Japan}
\newcommand{\spin}{Center for Spintronics Research Network, Keio University, Yokohama, Kanagawa 223-8522, Japan}
\newcommand{\RCEC}{Research Center for Emerging Computing Technologies, National Institute of Advanced Industrial Science and Technology (AIST), Tsukuba Ibaraki 305-8568, Japan}
\newcommand{\NECAIST}{NEC-AIST Quantum Technology Cooperative Research Laboratory, National Institute of Advanced Industrial Science and Technology (AIST), Tsukuba, Ibaraki 305-8568, Japan}
\newcommand{\nanomate}{Nanomaterials Research Institute, Kanazawa University, Kanazawa, Ishikawa 920-1192, Japan}
\newcommand{\AIST}{National institute of Advanced Industrial Science and Technology (AIST), Tsukuba, Ibaraki 305-8568, Japan}
\newcommand{\ICR}{Institute for Chemical Research, Kyoto University, Uji, Kyoto 611-0011, Japan}
\newcommand{\csrnkyoto}{Center for Spintronics Research Network, Kyoto University, Uji, Kyoto 611-0011, Japan}
\newcommand{\ICQMS}{International Center for Quantum-field Measurement Systems for Studies of the Universe and Particles (QUP), KEK, Tsukuba, Ibaraki 305-0801, Japan}
\newcommand{\utokyo}{Department of Physics, The University of Tokyo, Bunkyo-ku, Tokyo 113-0033, Japan}
\newcommand{\IPI}{Institute for Physics of Intelligence, The University of Tokyo, Bunkyo-ku, Tokyo 113-0033, Japan}
\newcommand{\TQSI}{Trans-scale Quantum Science Institute, The University of Tokyo, Bunkyo-ku, Tokyo 113-0033, Japan}
\makeatletter
\def\@email#1#2{%
 \endgroup
 \patchcmd{\titleblock@produce}
  {\frontmatter@RRAPformat}
  {\frontmatter@RRAPformat{\produce@RRAP{*#1\href{mailto:#2}{#2}}}\frontmatter@RRAPformat}
  {}{}
}%
\makeatother

\usepackage[]{hyperref}
\usepackage{xcolor}
\definecolor{AIPBlue}{RGB}{61, 180, 229}
\hypersetup{
    colorlinks,
    linkcolor=AIPBlue,
    citecolor=AIPBlue,
    urlcolor=AIPBlue
}

\begin{document}

\preprint{AIP/123-QED}

\title[Sample title]{Frequency-tunable magnetic field sensing using
 continuous-wave optically detected magnetic resonance with nitrogen-vacancy centers in diamond}
\author{Ryusei Okaniwa}
\affiliation{\keio}
\affiliation{\spin}
 \altaffiliation{}

\author{Takumi Mikawa}
\affiliation{\keio}
\affiliation{\spin}

\author{Yuichiro Matsuzaki}
\email[Author to whom correspondence should be addressed: ]{ymatsuzaki872@g.chuo-u.ac.jp}
\affiliation{\RCEC}
\affiliation{\NECAIST}

\author{Tatsuma Yamaguchi}
\affiliation{\keio}

\author{Rui Suzuki}
\affiliation{\keio}
\affiliation{\spin}
 
\author{Norio Tokuda}
\affiliation{\nanomate}
\author{Hideyuki Watanabe}
\affiliation{\AIST}
\author{Norikazu Mizuochi}
\affiliation{\ICR}
\affiliation{\csrnkyoto}
\affiliation{\ICQMS}
\author{Kento Sasaki}
\affiliation{\utokyo}
\author{Kensuke Kobayashi}
\affiliation{\utokyo}
\affiliation{\IPI}
\affiliation{\TQSI}
\author{Junko Ishi-Hayase}
\affiliation{\keio}
\affiliation{\spin}
\email[Author to whom correspondence should be addressed: ]{hayase@appi.keio.ac.jp}

\date{\today}

\begin{abstract}
The nitrogen-vacancy (NV) center is a promising candidate to realize practical quantum sensors with high sensitivity and high spatial resolution, even at room temperature and atmospheric pressure. In conventional high-frequency AC magnetometry with NV centers, the setup requires a pulse sequence with an appropriate time synchronization and strong microwave power. To avoid these practical difficulties, AC magnetic field sensing using continuous-wave optically detected magnetic resonance (CW-ODMR) was recently demonstrated. That previous study utilized radio frequency (RF) dressed states generated by the coherent interaction between the electron spin of the NV center and the RF wave. However, the drawback of this method is that the detectable frequency of the AC magnetic fields is fixed. Here, we propose and demonstrate frequency-tunable magnetic field sensing based on CW-ODMR. In the new sensing scheme, we obtain RF double-dressed states by irradiation with a RF field at two different frequencies. One creates the RF dressed states and changes the frequency of the target AC field. The other is a target AC field that induces a change in the CW-ODMR spectrum by generating the RF double-dressed states through coherent interaction with the RF dressed states. The sensitivity of our method is estimated to be comparable to or even higher than that of the conventional method based on a RF field with a single frequency. The estimated bandwidth is 7.45~MHz, higher than that of the conventional method using the RF dressed states. Our frequency-tunable magnetic field sensor based on CW-ODMR paves the way for new applications in diamond devices.
\end{abstract}

\maketitle

\section{\label{sec:introduction}INTRODUCTION}
In natural science, quantum sensing has received considerable attention because it can facilitate the investigation of physical properties on the nanometer scale and the evaluation of miniaturized devices. In recent years, the nitrogen-vacancy (NV) center in diamond has emerged as a promising candidate for a  magnetometry with high spatial resolution. The significant characteristic of the NV center is long coherence time, even at room temperature and atmospheric pressure, \cite{Balasubramanian2009,Herbschleb2019} unlike the other quantum sensors such as SQUIDs. \cite{Halbertal2016,Vasyukov2013}
In addition, the electron spin states can be initialized by irradiation with a green laser\cite{Harrison2004} and read out through the spin-dependent photoluminescence. \cite{Gruber1997,Jelezko2002,schirhagl2014nitrogen} Moreover, we can manipulate spin states in the NV center using microwave (MW) field.\cite{Jelezko2004} For practical magnetic field sensing using NV centers, many methods such as wide-field imaging,\cite{Pham2011,LeSage2013,Mizuno2020} AFM-based methods, \cite{balasubramanian2008nanoscale,Degen2008,huxter2022scanning} and vector magnetic field sensing\cite{Maertz2010,Steinert2010,Pham2011,Kitazawa2017,Yahata2019,wang2021nanoscale} have been proposed and demonstrated.

To realize sensitive AC magnetic field detection with NV centers, pulsed measurement such as spin echo is typically used.\cite{Taylor2008,maze2008nanoscale,pham2012enhanced,loretz2013radio,Wolf2015a,stark2017narrow} While such pulsed measurement is advantageous for improving sensitivity because of the extended coherence time ($T_2$),\cite{bar2012suppression,bar2013solid} it requires a sophisticated setup to generate pulse sequences with appropriate synchronization. In addition, when the pulsed-measurement is used in CCD-based wide-field imaging, we must deal with significantly different time scales for pulse sequence and imaging frame rates.
It is also vital to improve the pulse precision, as it is challenging to apply spatially uniform and strong MW over a large area.

To avoid such difficulties, some of the present authors proposed and demonstrated sensing AC fields with MHz frequencies (RF fields) with MHz frequencies via continuous-wave optically detected magnetic resonance (CW-ODMR).\cite{Saijo2018,Yamaguchi2019} This technique utilizes not only a MW, but also a RF field, which is the target field. Due to the coherent interaction between the electron spin states and the driving fields, RF dressed states are created. As a result, we observe a change in the CW-ODMR signal. In this technique, CW-ODMR is used to measure the spectrum of the RF-dressed states created by the coherent interaction of electron spins with the target RF magnetic field. The RF field strength can be estimated from the resulting ODMR spectrum. In this approach, a sophisticated setup and strong MW power are not require, unlike the conventional pulsed measurement. However, the target frequency is fixed at the non-adjustable physical properties of the sensor, and any detuning from this value rapidly reduces the sensitivity. This narrow band property limits the broad application of this ambitious technique.

Here, we propose and demonstrate frequency-tunable version of the CW-ODMR based AC magnetic field sensing. One is the control RF ($\mathrm{RF_{control}}$) field that generates RF dressed states of the NV centers in the diamond. The other is the target RF ($\mathrm{RF_{target}}$) field, which is the AC magnetic field whose amplitude is to be estimated by our method. The coherent interaction between the RF dressed states and the $\mathrm{RF_{target}}$ field generates RF double-dressed states and induces measurable signals during CW-ODMR measurement. In addition, since the energy levels of the RF dressed states can be controlled by changing the $\mathrm{RF_{control}}$ field power, the detectable frequency can be tuned. Our proof of principle experiments proves that the detection bandwidth can be increased compared with those of the previously reported methods. Our results open the door to realizing AC magnetic field sensor with a large bandwidth by using CW-ODMR on the NV centers.

\section{THEORY}
We describe the theory underlying our proposed method. We apply a DC magnetic field perpendicular to a specific axis of the NV center and define this direction as the $x$ axis. In this case, the eigenstates of the NV centers are approximately described by the so-called bright state or dark state. Importantly, the RF field can induce transitions between these states. In addition, by applying a perpendicular DC magnetic field, we can suppress magnetic field noise.
 
The Hamiltonian (\(\hat{H}_{\mathrm{NV}}\)) of the NV center without the MW and RF fields is described as
follows: \cite{Yamaguchi2019,Yamaguchi2020,tabuchi2023temperature} 
\begin{eqnarray}
\label{eq:HNV}
\hat{H}_{\rm{NV}}&&\simeq D'\hat{S}_z^2 +E_x'\left(\hat{S}_x^2-\hat{S}_y^2\right),\\
D'&&=D+\frac{3}{2}\frac{({\gamma}_eB_x)^2}{D+E_x},\\
E_x'&&=E_x+\frac{1}{2}\frac{({\gamma}_eB_x)^2}{D + E_x}\label{eq:E'x},
\end{eqnarray}
where \(\hat{\bm{S}} = (\hat{S_x}, \hat{S_y}, \hat{S_z})\) is the spin-1 operator of the electron spin, \(D/2\pi (\simeq 2.87\)~GHz) is the zero-field splitting, ${ E }_{ x }$ is the strain along the $x$ direction, \(\gamma_e/2\pi (\simeq 28\)~GHz/T) is the gyromagnetic ratio of the electron spin, and $B_x$ is the amplitude of the DC magnetic field. The eigenstates of Eq.~(\ref{eq:HNV}) are described as \(|0\rangle \), \(|B\rangle =\frac{1}{\sqrt{2}}(|+1\rangle + |-1\rangle)\) and \(|D\rangle=\frac{1}{\sqrt{2}}(|+1\rangle-|-1\rangle)\), where \(|+1\rangle, |-1\rangle, |0\rangle\) are distinguished by their magnetic quantum numbers. \(|B\rangle\) (\(|D\rangle\)) is the so-called bright (dark) state. 

\begin{figure}[ht]
\centering
\includegraphics[]{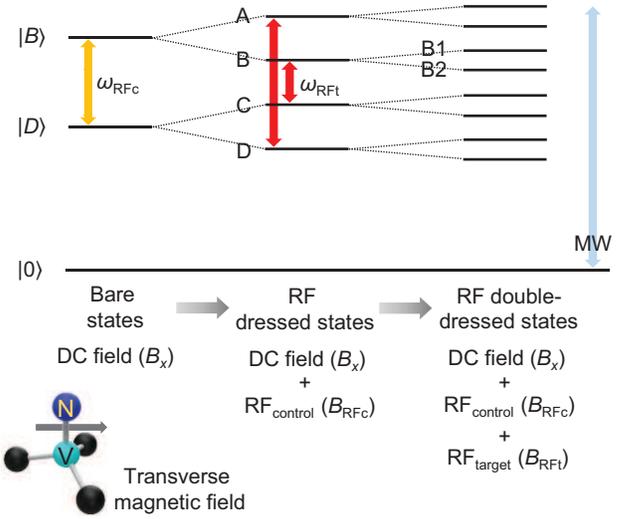}
\caption{Energy levels of NV centers in our setup. In bare states, the spin triplet is composed of \(|B\rangle\) (bright state), \(|D\rangle\) (dark state), and \(|0\rangle\). Irradiation with the  $\mathrm{RF_{control}}$ field, which is resonant between \(|B\rangle\) and \(|D\rangle\), generates the RF dressed states. In this situation, four dips are observed by CW-ODMR. Note that the energy levels of the RF dressed states can be controlled by the  $\mathrm{RF_{control}}$ field amplitude. Moreover, we apply the  $\mathrm{RF_{target}}$ field as the target AC magnetic field. When the $\mathrm{RF_{target}}$ field is resonant between the RF dressed states, additional splitting occurs, and eight photoluminescence dips are finally observed in the CW-ODMR measurement, corresponding to the RF double-dressed states. In our proposed scheme, the energy splitting in the RF double-dressed states is utilized for CW-ODMR-based sensing.}
\abovecaptionskip=0pt
\label{NVstate}
\end{figure}

We calculate the dynamics of the NV center under irradiation by the  MW and two RF fields, as shown in Fig.~\ref{NVstate}. The MW along the $x$ ($y$) direction induces the transition between \(|0\rangle\) and \(|B\rangle\) (\(|D\rangle\)). Whichever MW direction is considered, we go through the same calculation process. The first RF field ($\mathrm{RF_{control}}$) is resonant between \(|B\rangle\) and \(|D\rangle\) to generate the RF dressed states, while the second RF field ($\mathrm{RF_{target}}$) is the target AC magnetic field. We use the MW to probe the interaction between the RF dressed states and the $\mathrm{RF_{target}}$. The total Hamiltonian ($H_{\mathrm{tot}}$) of the NV centers with $\mathrm{RF_{control}}$, $\mathrm{RF_{target}}$, and MW is given as

\begin{align*}
\hat{H}_{\mathrm{tot}}\simeq & D'\hat{S}_z^2 + E_x'(\hat{S}_x^2 -\hat{S}_y^2)\\
&+{\gamma}_eB_{\mathrm{RFc}} \hat{S}_z\cos (\omega_{\mathrm{RFc}}t)\\
&+{\gamma}_eB_{\mathrm{RFt}} \hat{S}_z
\cos (\omega_{\mathrm{RFt}}t)   \\
&+{\gamma}_eB_{\mathrm{MW}} \hat{S}_x \cos (\omega_{\mathrm{MW}} t),
\stepcounter{equation}\tag{\theequation} 
\end{align*}
where, $B_{\mathrm{RFc}}$, $B_{\mathrm{RFt}}$, and $B_{\mathrm{MW}}$ are the amplitudes of the $\mathrm{RF_{control}}$ field, the $\mathrm{RF_{target}}$ field, and the MW, respectively, while $\omega_{\mathrm{RFc}}$, $\omega_{\mathrm{RFt}}$, and $\omega_{\mathrm{MW}}$ are the frequencies of the $\mathrm{RF_{control}}$ field, the $\mathrm{RF_{target}}$ field, and the MW, respectively. Here, the terms about the RF fields along the $x$ and $y$ direction are dropped due to the rotating wave approximation (RWA). In a rotating frame with $\hat{U}^{(1)}=e^{-i(\omega_{\mathrm{RFc}}|D\rangle \langle D|+\omega_{\mathrm{MW}}|0\rangle \langle 0|)t}$, the effective Hamiltonian ($\hat{H}^{(1)}_{{\mathrm{tot}}}$) is given as
\begin{align*}
\hat{H}^{(1)}_{{\mathrm{tot}}}\simeq &(D'+E_x'-\omega_{\mathrm{MW}})|B\rangle \langle B|\\
&+(D'-E_x'+\omega_{\mathrm{RFc}}-\omega_{\mathrm{MW}})|D\rangle \langle D|\\
&+\frac{\gamma_e B_{\mathrm{RFc}}}{2} (|B\rangle \langle D|-|D\rangle \langle B|) \\
&+\frac{\gamma_e B_{\mathrm{MW}} }{2} (|B\rangle \langle 0|+|0\rangle \langle B|)\\
&+\frac{\gamma_e B_{\mathrm{RFt}}}{2}(|B\rangle \langle D|e^{i(\omega_{\mathrm{RFc}} -\omega_{\mathrm{RFt}})t}\\
&+|D\rangle \langle B|e^{-i(\omega_{\mathrm{RFc}} -\omega_{\mathrm{RFt}})t}),
\stepcounter{equation}\tag{\theequation}
\end{align*}
where we have used the RWA for the MW and $\mathrm{RF_{control}}$ fields. Since we apply the resonant $\mathrm{RF_{control}}$ ($\omega_{\mathrm{RFc}}=2E_x'$), 
we can simplify Eq.~(\ref{eq:afterRWA1simp}) as follows:
\begin{align*}
\hat{H}^{(1)}_{{\mathrm{tot}}} \simeq &(D'+E_x'-\omega_{\mathrm{MW}}+\frac{\gamma_eB_{\mathrm{RFc}} }{2})|+1\rangle \langle +1|\\
&+(D'+E_x'-\omega_{\mathrm{MW}}-\frac{\gamma_e B_{\mathrm{RFc}} }{2})|-1\rangle \langle -1|\\
&+\frac{\gamma_e B_{\mathrm{MW}} }{2} (|B\rangle \langle 0|+|0\rangle \langle B|)\\
&+\frac{\gamma_eB_{\mathrm{RFt}}}{2}(|B\rangle \langle D|e^{i(2E_x' -\omega_{\mathrm{RFt}})t}\\
&+|D\rangle \langle B|e^{-i(2E_x' -\omega_{\mathrm{RFt}})t}).
\stepcounter{equation}\tag{\theequation}
\label{eq:afterRWA1simp}
\end{align*}

Furthermore, we define another rotating frame as
$\hat{U}^{(2+)}=e^{-i(2E_x' -\omega_{\mathrm{RFt}})|-1\rangle \langle-1|t}$
($\hat{U}^{(2-)}=e^{i(2E_x' -\omega_{\mathrm{RFt}})|+1\rangle \langle +1|t}$).
By using the RWA for the $\mathrm{RF_{target}}$ field, we obtain a Hamiltonian $\hat{H}^{(2+)}_{\mathrm{tot}}$ ($\hat{H}^{(2-)}_{\mathrm{tot}}$) as follows:

\begin{align*}
\hat{H}^{(2\pm)}_{{\mathrm{tot}}} \simeq &(D'+E_x'-\omega_{\mathrm{MW}}+\frac{\gamma_e B_{\mathrm{RFc}}}{2})|+1\rangle \langle+1|\\
&+(D'+E_x'-\omega_{\mathrm{MW}}-\frac{\gamma_eB_{\mathrm{RFc}}}{2}) |-1\rangle \langle -1|\\
&\pm (2E_x'-\omega_{\mathrm{RFt}})|\mp 1\rangle \langle \mp 1| \\
&+\frac{\gamma_e B_{\mathrm{MW}} }{2\sqrt{2}} (|\pm 1\rangle \langle 0|+|0\rangle \langle \pm 1|)\\
& +\frac{\gamma_eB_{\mathrm{RFt}} }{4} (|+1\rangle \langle -1|+|-1\rangle \langle +1|).
\stepcounter{equation}\tag{\theequation}
\label{eq:afterRWA2}
\end{align*}
To be consistent with the theory, throughout of this paper, we apply the weak MW fields in the experiment. We can treat this system as harmonic oscillators.\cite{diniz2011strongly,zhu2014observation,matsuzaki2015improving,Yamaguchi2019}
By defining \(\hat{b}^{\dagger} = |+1\rangle \langle 0|\) 
\((\hat{d}^{\dagger}=|-1\rangle \langle 0|)\) as a creation operator and \(\hat{b} = |0\rangle \langle +1|\) \((\hat{d}=|0\rangle \langle -1|)\) as an annihilation operator, we can adopt Eq.~(\ref{eq:afterRWA2}) and rewrite {$\hat{U}^{(2+)}=e^{-i(2E_x' -\omega_{\mathrm{RFt}})|-1\rangle \langle-1|t}$} as
\begin{align*}
\hat{H}^{(2+)}_{{\mathrm{tot}}} \simeq \omega _ { b } \hat { b } ^ { \dagger } \hat { b } 
+\omega _ { d } \hat { d } ^ { \dagger } \hat { d } 
+\lambda \left( \hat { b } + \hat { b } ^ { \dagger }  \right)
+J\left(\hat { b } \hat { d } ^ { \dagger }+\hat { b } ^ { \dagger } \hat { d } \right),
\label{eq:afterRWA2-2}
\stepcounter{equation}\tag{\theequation}
\end{align*}
where $\omega _ { b }=D'+E_x'-\omega_{\mathrm{MW}}+\frac{\gamma_e B_{\mathrm{RFc}}}{2}$, $\omega _ { d }=D'+E_x'-\omega_{\mathrm{MW}}-\frac{\gamma_e B_{\mathrm{RFc}}}{2}+(2E_x' -\omega_{\mathrm{RFt}})$, $\lambda=\frac{\gamma_e B_{\mathrm{MW}}}{2\sqrt{2}}$, and $J= \frac{\gamma_e B_{\mathrm{RFt}}}{4}$. 
From this Hamiltonian, we can write the Heisenberg equations and consider a steady state of the system. In this case, we calculate the probability $p_0$ that the system is in \(|0\rangle\) as follows: \cite{zhu2014observation,matsuzaki2016optically}
\begin{eqnarray}
 p_0 &&=1-p_b -p_d, \label{eq:p0}\\
 p_b~
 &&= |\frac { - \lambda _ { b } \left( \omega _ { d } - \mathrm{i} \Gamma _ { d } \right) } { \left( \omega _ { b } - \mathrm{i} \Gamma _ { b } \right) \left( \omega _ { d } - i \Gamma _ { d } \right) - J ^ { 2 } }|^2,  \\ 
p_d~ &&= |\frac { \lambda _ { b } J } { \left( \omega _ { b } - \mathrm{i} \Gamma _ { b } \right) \left( \omega _ { d } - \mathrm{i} \Gamma _ { d } \right) - J ^ { 2 } }|^2\label{eq:d},
\end{eqnarray}
 where  \(\Gamma_b\) (\(\Gamma_d\)) is an effective decay rate of the bright (dark) state.

By using Eqs.~(\ref{eq:p0})-(\ref{eq:d}), we can calculate the resonant frequency of the MW field ($\omega_{\mathrm{res1\pm}}$ and $\omega_{\mathrm{res2\pm}}$) as
\begin{align*}
\omega_{\mathrm{res1\pm}}= &D'+2E_x'-\frac{\omega_{\mathrm{RFt}}}{2}\\
&\pm\frac{1}{4}\sqrt{4(-2E_x'+\omega_{\mathrm{RFt}}+\gamma_e{B_{\mathrm{RFc}}})^2+(\gamma_e{B_{\mathrm{RFt}})^2}},
\stepcounter{equation}\tag{\theequation}
\label{lowans1}
\end{align*}
\begin{align*}
\omega_{\mathrm{res2\pm}}=&D'+\frac{\omega_{\mathrm{RFt}}}{2}\\
&\pm\frac{1}{4}\sqrt{4(-2E_x'+\omega_{\mathrm{RFt}}+\gamma_e{B_{\mathrm{RFc}}})^2+(\gamma_e{B_{\mathrm{RFt}})^2}}.
\stepcounter{equation}
\tag{\theequation}
\label{lowans2}
\end{align*}
From these equations, we obtain
$\omega_{\mathrm{RFt}}\simeq 2E'_{x} - \gamma_{e}B_\mathrm{{RFc}}$ as a condition that the target RF field is resonant with the dressed states, as shown in Fig.~\ref{NVstate}. When this condition is satisfied, we expect to observe a dip in the CW-ODMR at the MW frequency of Eq.~\eqref{lowans1} and Eq.~\eqref{lowans2}. In addition, the resonant frequency in Eq.~\eqref{lowans1} is directly affected by the value of $E_x'$, indicating that this resonance would be sensitive to electric field noise. Thus, to improve the sensitivity, we do not use this resonance in our method. On the other hand, in Eq.~\eqref{lowans2}, we can suppress the effect of $E_x'$ by increasing the amplitude of the control field, which should improve the sensitivity.

\begin{figure*}[htb]
\centering
\includegraphics[]{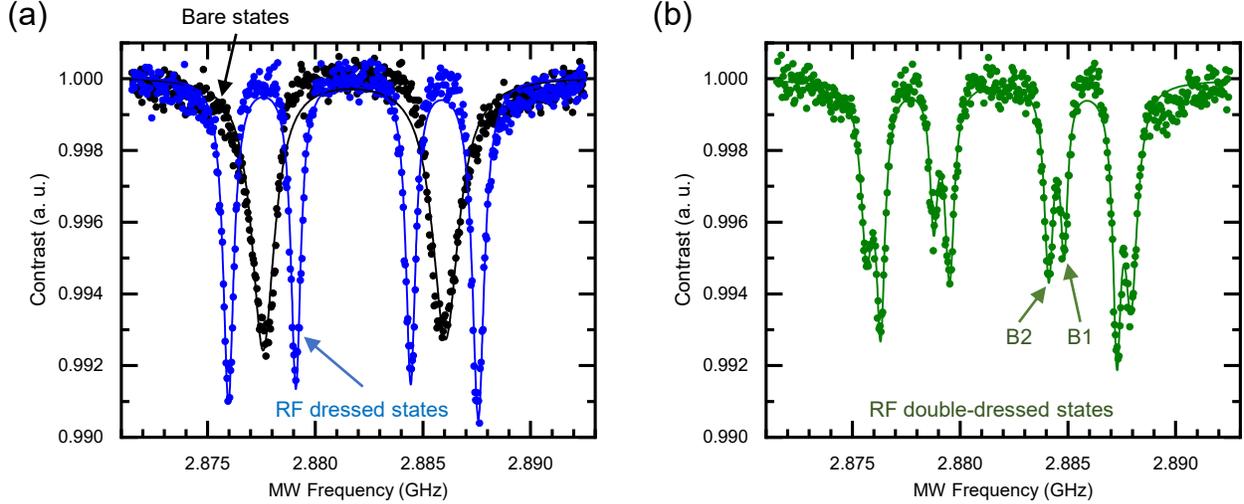}
\caption{(a) CW-ODMR spectra under a perpendicular magnetic field with/without the $\mathrm{RF_{control}}$ field of 8.47~MHz (101~$\mu$T). The black line corresponds to the bare states, whereas the blue line corresponds to the RF dressed states. (b) CW-ODMR spectrum (green line) with the $\mathrm{RF_{target}}$ field of 5.34~MHz (29.5~$\mu$T) in addition to the $\mathrm{RF_{control}}$ field (8.47~MHz, 101~$\mu$T). Due to the creation of the RF double-dressed states, eight dips can be observed.}
\abovecaptionskip=10pt
\label{f:ODMRwith/outRF}
\end{figure*}

We can adopt $\hat{U}{'}^{(2+)}=e^{i(2E_x' -\omega_{\mathrm{RFt}})|-1\rangle \langle-1|t}$ or $\hat{U}{'}^{(2-)}=e^{-i(2E_x' -\omega_{\mathrm{RFt}})|+1\rangle \langle+1|t}$ instead of $\hat{U}^{(2\pm)}$. In this case, by performing similar calculations, we obtain the resonant frequency of the MW field ($\omega_{\mathrm{res3\pm}}$ and $\omega_{\mathrm{res4\pm}}$) as
\begin{align*}
\omega_{\mathrm{res3\pm}}=&D'+\frac{\omega_{\mathrm{RFt}}}{2}\\
&\pm\frac{1}{4}\sqrt{4(2E_x'-\omega_{\mathrm{RFt}}+\gamma_eB_{\mathrm{RFc}})^2+(\gamma_e{B_{\mathrm{RFt}})^2}},
\stepcounter{equation}\tag{\theequation}
\label{highans1}
\end{align*}
\begin{align*}
\omega_{\mathrm{res4\pm}}=&D'+2E_x'-\frac{\omega_{\mathrm{RFt}}}{2}\\
&\pm\frac{1}{4}\sqrt{4(2E_x'-\omega_{\mathrm{RFt}}+\gamma_e{B_{\mathrm{RFc}}})^2+(\gamma_e{B_{\mathrm{RFt}})^2}}.
\stepcounter{equation}\tag{\theequation}
\label{highans2}
\end{align*}
From these equations, we obtain a resonant condition of the RF as $\omega_{\mathrm{RFt}}\simeq 2E'_{x} + \gamma_{e}B_\mathrm{{RFc}}$ (as shown in Fig.~\ref{NVstate}). Again, when this condition is satisfied, we expect to observe a dip in the CW-ODMR at the MW frequency of Eq.~\eqref{highans1} and Eq.~\eqref{highans2}. The resonant frequency in Eq.~\eqref{highans2} is affected by the value of $E_x'$; thus, we do not use this resonance to avoid the effect of electric field noise in our method.

These analytical solutions indicate that the resonant frequency of CW-ODMR changes linearly with the $\mathrm{RF_{target}}$ field amplitude when the frequency of the $\mathrm{RF_{target}}$ field is equal to \(2E'_x \pm \gamma_e {B_{\mathrm{RFc}}}\), as shown in Fig.~\ref{NVstate}. In addition, the intensity of the $\mathrm{RF_{target}}$ field can be measured from the changes in the resonant frequency of CW-ODMR. Since we can measure the $\mathrm{RF_{target}}$ field at the frequency of $2E'_x \pm \gamma_e {B_{\mathrm{RFc}}}$, we can determine the detectable frequency by changing the $\mathrm{RF_{control}}$ field amplitude. Although we assume that we drive the NV centers using the MW along the $x$ direction, we can perform similar calculations with the MW aligned along the $y$ direction and obtain the same number of resonances.

\section{EXPERIMENTAL RESULTS}
\subsection{Setup and CW-ODMR measurement}
We describe the setup of our experiments. We use the same diamond sample used in Ref.~\onlinecite{tabuchi2023temperature}. The ensemble of NV centers has one axis oriented along the (111) direction. The measurement setup is basically the same as that in Refs.~\onlinecite{Yamaguchi2019} and \onlinecite{tabuchi2023temperature}, except that we apply the second RF field to the NV centers. The sample of ensemble NV centers was placed on the MW antenna.\cite{Sasaki2016} perpendicular DC magnetic field was applied by a permanent magnet below the MW antenna. The $\mathrm{RF_{control}}$ field and the $\mathrm{RF_{target}}$  field were irradiated from a single copper wire that was closely attached to the sample surface.

\begin{figure*}[hbt]
\begin{center}
\includegraphics[]{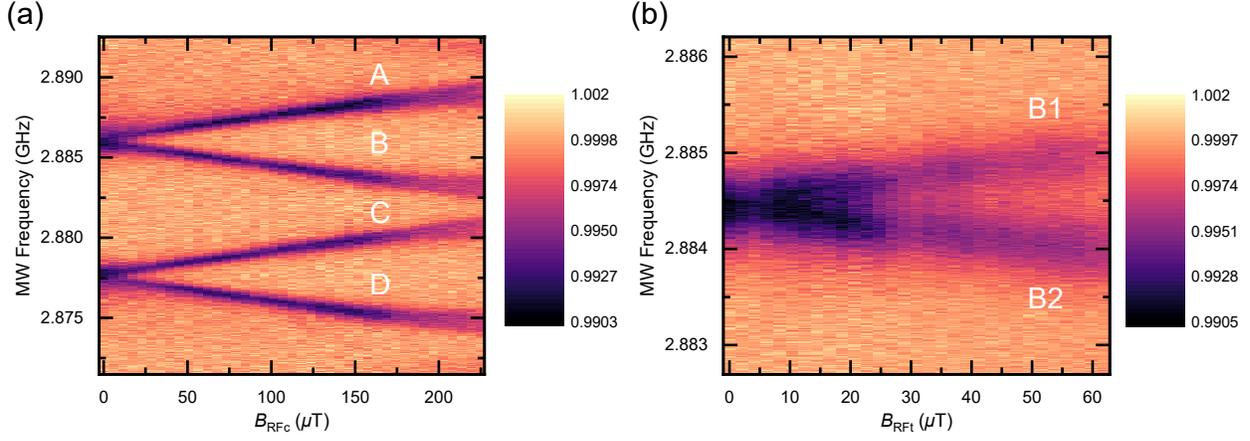}
\caption{(a) Color mapping of the CW-ODMR spectra obtained by sweeping the MW frequency and the $\mathrm{RF_{control}}$ field amplitude of 8.47~MHz. A darker color area indicates a photoluminescence dip where a resonant MW frequency can be found. The four dips split linearly according to increase in the $\mathrm{RF_{control}}$ field amplitude. (b) Color mapping of the CW-ODMR spectra obtained by sweeping the MW frequency and the $\mathrm{RF_{control}}$ field amplitude of 5.34~MHz. Note that we extracted one dip in the RF dressed states. In addition, the  $\mathrm{RF_{control}}$ field is applied at 8.47~MHz and 101~$\mu$T. Similar to the result shown in Fig.~3(a), linear energy splitting can be observed.}
\abovecaptionskip=10pt
\label{f:ODMRcolormappping}
\end{center}
\end{figure*}

Before demonstrating CW-ODMR for AC magnetic field sensing, we needed to perform CW-ODMR for calibration. 

First, we performed the CW-ODMR measurement by sweeping the MW frequency under a perpendicular magnetic field, as shown in 
Fig.~\ref{f:ODMRwith/outRF}(a). We observed two dips, as shown by the black line in Fig.~\ref{f:ODMRwith/outRF}(a). One of them corresponds to the transition from \(|0\rangle\) to \(|B\rangle\), while the other corresponds to the transition from \(|0\rangle\) to \(|D\rangle\). We found that the resonant frequencies of \(|B\rangle\) and \(|D\rangle\) are 2.878~GHz and 2.886~GHz, respectively. Their difference 8.47~MHz corresponds to the resonance frequency of $\mathrm{RF_{target}}$ for creating the RF dressed state.

Second, we performed the CW-ODMR measurement under irradiation of the $\mathrm{RF_{control}}$ field with a frequency of 8.47~MHz by sweeping the MW frequency. The results are shown by the blue line in Fig.~\ref{f:ODMRwith/outRF}(a). We observed four dips and confirmed the generation of the RF dressed states. We also found that the linewidth of the RF dressed states was narrower than that of the bare states without RF fields (see Eq.~\eqref{eq:doublereso1} and Eq.~\eqref{eq:doublereso2} ). This is consistent with the fact that the RF-dressed states are more robust against the electric noise than the bare states.\cite{Yamaguchi2019,tabuchi2023temperature}

Third, we performed the CW-ODMR measurement under irradiation by the $\mathrm{RF_{target}}$ ($\mathrm{RF_{control}}$) field with a frequency of 5.34 (8.47)~MHz; this frequency corresponds to 
$2E'_{x}-\gamma_e B_{\mathrm{RF_c}}$ ($2E'_{x}$). The results are shown by the green line in Fig.~\ref{f:ODMRwith/outRF}(b). We observed eight dips, which come from the RF double-dressed states, as we show in Eqs.~(\ref{lowans1})-(\ref{lowans2}). 

Fourth, we performed the CW-ODMR measurement by sweeping the amplitude of the $\mathrm{RF_{control}}$ field (with a frequency of 8.47~MHz) and MW frequency. The color mapping of the CW-ODMR spectra (four dips) in the RF dressed states is shown in Fig.~\ref{f:ODMRcolormappping}(a). As in previous studies \cite{Saijo2018,Yamaguchi2019}, the splitting became larger as the $\mathrm{RF_{control}}$ field amplitude increased. Here, we succeeded in controlling the energy splitting of the RF dressed states by changing the $\mathrm{RF_{control}}$ field amplitude. This means that we can tune the energy splitting to be resonant with the target RF frequency by changing the $\mathrm{RF_{control}}$ field amplitude. In addition, the proportionality constant between the $\mathrm{RF_{control}}$ field amplitude and energy splitting is approximately 25.7~GHz/T. If we assume that the DC magnetic field is applied perpendicular to the NV axis, the proportionality constant can be calculated as 28~GHz/T, which is different from the observed value. We expect that this difference comes from the misalignment of the DC magnetic fields.

Finally, we performed the CW-ODMR with the $\mathrm{RF_{control}}$ field at a frequency of 8.47~MHz and an amplitude of 101~$\mu\mathrm{T}$ while we swept both the MW frequency and the $\mathrm{RF_{target}}$ field amplitude with a frequency of 5.34~MHz.
The results are shown in Fig.~\ref{f:ODMRcolormappping}(b). Here, we focus on one (labeled "B" in Fig.~\ref{NVstate}) of the four dips in the RF dressed states; this dip is split into two dips (labeled "B1" and "B2" in Fig.~\ref{NVstate}) when we apply the $\mathrm{RF_{target}}$ field, as shown in Eq.~(\ref{lowans2}). During CW-ODMR, we observe splitting due to the RF double-dressed states. This splitting becomes larger as we increase the $\mathrm{RF_{target}}$ field amplitude. In addition, the proportionality constant between the $\mathrm{RF_{target}}$ field amplitude and energy splitting is approximately 13.7~GHz/T, smaller than the theoretically predicted value of 14~GHz/T. We expect that this difference also comes from the misalignment of the DC magnetic fields, as discussed above.

\subsection{Sensitivity and frequency tunability}

\begin{figure}[hbt]
\begin{center}
\includegraphics[]{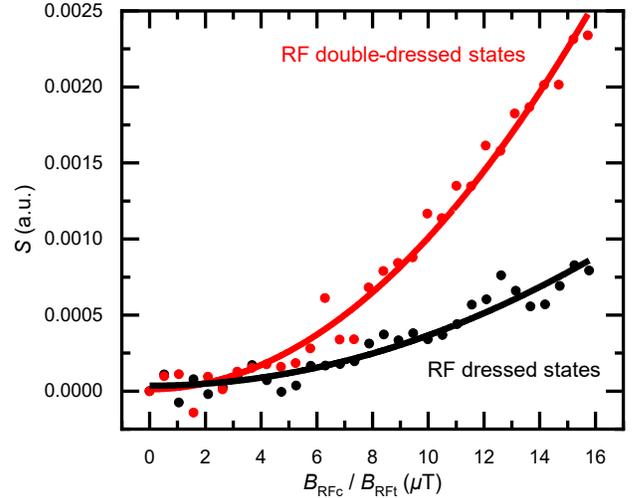}
\caption{Contrast change for a RF field of 7.14~MHz and MW frequency of 2.885~GHz in CW-ODMR. The black dots and red dots correspond to signal changes derived from the $\mathrm{RF_{control}}$ field and  $\mathrm{RF_{target}}$ field. Both dots are fitted with a quadratic function (\(S = aB_{\mathrm{RFt(c)}}^2 + S_0\)), which is needed to calculate the sensitivity. Moreover, the contrast change in the new scheme is larger than that in our previous scheme.}
\abovecaptionskip=10pt
\label{f:ODMRsensingmethod}
\end{center}
\end{figure}

We can estimate the amplitude of $\mathrm{RF_{target}}$ from the change in the contrast of the CW-ODMR as follows. When the $\mathrm{RF_{target}}$ field is applied with a frequency of $\omega_{\mathrm{RFt}}=2E'_x -\gamma_e B_{\mathrm{RFc}}$ ($\omega_{\mathrm{RFt}}=2E'_x +\gamma_e B_{\mathrm{RFc}}$), we fix the MW frequency at $D'+E'_x-\frac{1}{2}\gamma_e B_{\mathrm{RFc}}$ ($\omega_{\mathrm{MW}} = D'+E'_x+\frac{1}{2}\gamma_e B_{\mathrm{RFc}}$) and measure the change in the CW-ODMR contrast. Below, we explain the sensitivity and frequency tunability of this method.

We measured the amplitude of CW-ODMR signal when we applied the target field. More specifically, we set the target frequency to 7.14~MHz. The results obtained using our approach and the previous scheme are compared in Fig.~\ref{f:ODMRsensingmethod}. The vertical axis $S$ is the change of the signal amplitude from the $B_{\mathrm{RFt(c)}} = 0$. In both schemes, it quadratically with the amplitude of the target RF field.\cite{Saijo2018,Yamaguchi2019} The obtained signal is significantly larger than that of the previous method. This is because in the previous method, the resonant frequency is detuned by 1.33~MHz with respect to the target frequency; in contrast, in the new scheme, we can use the control RF field to tune the resonant frequency to match the target frequency. Here, it is worth mentioning that in previous scheme where only one RF field is used, we defined ${B_{\mathrm{RFc}}}$ as the sensing target.

Finally, we experimentally estimated the sensitivity of our method and compare it with that of the previous method. We review the previous method in Appendix \ref{appendix}. We measured $\delta S$, the signal fluctuation per second. More specifically, we collected the signal fluctuations over several measurement times using an avalanche photodiode and obtained $\delta S$ by linear fitting. We also estimated $\frac{\mathrm{d}S}{\mathrm{d}B_{\mathrm{RFt}}}$ ($\frac{\mathrm{d}S}{dB_{\mathrm{RFc}}}$) for the new (previous) method. From these estimates, we obtained the sensitivity as $\delta B_{\mathrm{RFt}}=\delta S/|\frac{\mathrm{d}S}{ dB_{\mathrm{RFt}}}|$ ($\delta B_{\mathrm{RFc}}=\delta S/|\frac{\mathrm{d}S}{dB_{\mathrm{RFc}}}|$) for our (previous) method. The sensitivity of the previous method is optimized around 8.47~MHz at the bandwidth of 3.75~MHz. However, the sensitivity rapidly decreases when the frequency is detuned from 8.47~MHz. In this case, the bandwidth is limited by the inhomogeneous linewidth. Compared with the previous scheme, our method has the advantage of an extended detectable frequency range. The optimal sensitivity is around 4.31 $\mu\mathrm{T}/\sqrt{\mathrm{Hz}}$ when the frequency is around $8$ MHz. As we increase or decrease the frequency, the sensitivity becomes worse. We define the bandwidth of the frequency when the sensitivity becomes half as the optimal one. The bandwidth is estimated to be 7.45~MHz, two times larger than that for the previous scheme. It is worth mentioning that due to electric field noise, we cannot measure the target field with our method when the frequency is in the range of 7.72--9.14~MHz. When the frequency is in this range, we can use the previous method. Thus, by combining our method and the previous method, we realize an AC magnetic field sensor with a large bandwidth.

\begin{figure}[hbt]
\begin{center}
\includegraphics[]{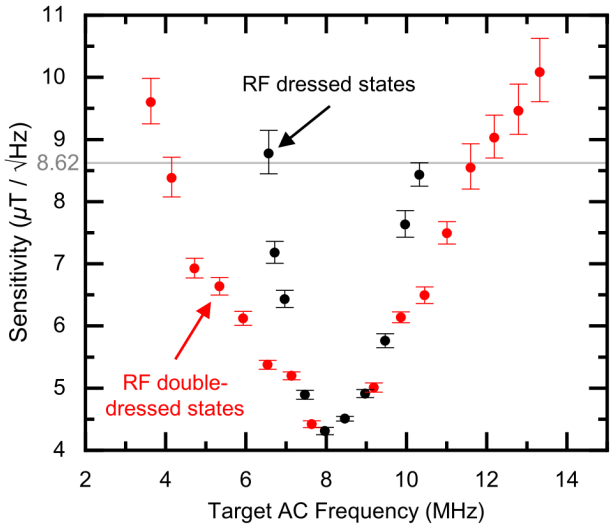}
\caption{The sensitivities for various target RF field frequencies obtained using our scheme (red dots) and the previous scheme (black dots).}
\label{f:sensing}
\end{center}
\end{figure}

We observed that the sensitivity of our scheme is comparable to that of the previous scheme, which can be explained as follows. According to the theoretical equations from Ref.~\onlinecite{Yamaguchi2019} and Eqs.~(\ref{lowans1})-(\ref{highans2}), the value of the energy splitting in the RF double-dressed states is half that in the RF dressed states. However, the creation of the RF dressed states suppresses environmental fluctuations such as electric noise. As a result, the linewidth of the photoluminescence dip is narrower, as shown in Fig.~\ref{f:ODMRwith/outRF}(a). This means that a larger contrast change can be obtained. Thus, this mechanism compensates the less responsiveness to the target RF field and contributes to the sensitivity shown in Fig.~\ref{f:sensing}.

\begin{figure}[hbt]
\begin{center}
\includegraphics[]{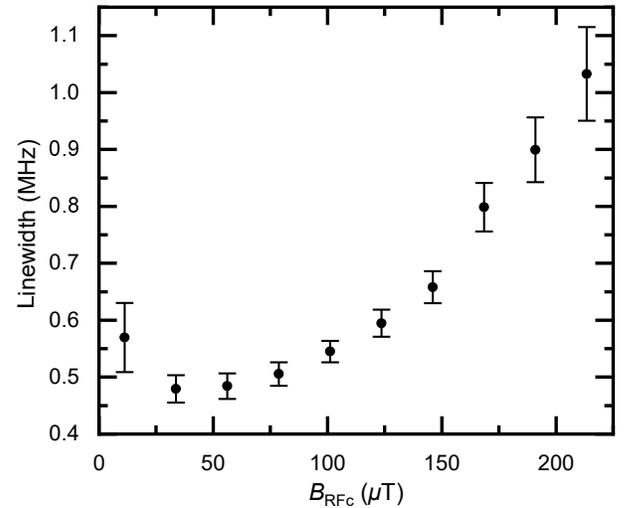}
\caption{The linewidth of the resonance of the RF-dressed states against the amplitude of $\mathrm{RF}_{\mathrm{control}}$. The linewidth is obtained from the Lorentzian fit. For small RF amplitudes, as we increase the RF amplitude, the linewidth becomes smaller because of the suprresion of the electric field noise. However, larger RF amplitudes lead to an increase of the the fluctuation of the RF, while the effect to suppress the electric field noise is saturated. There, in our case, for more than 33.7~$\mu$T, the linewidth monotonically increase with increasing the RF amplitude.} 
\label{f:linewidth}
\end{center}
\end{figure}

As we increase or decrease the frequency of the target field from 8.47~GHz, the sensitivity becomes worse. Here, we discuss possible reasons for this. To tune the frequency, we need to increase the amplitude of the control field. As discussed earlier, this control field can suppress the electrical noise. However, the fluctuation in the control field can also be a source of noise when the amplitude of the control field is large. This mechanism increases in CW-ODMR linewidth, as shown in Fig. \ref{f:linewidth}.
Such a fluctuation could cause the decrease in sensitivity when the frequency of the target field is detuned far from 8.47 MHz. A similar effect is discussed in Ref.~\onlinecite{cai2012robust}, in which concatenated dynamical decoupling was adopted.

The dependence on the detuning is assymetric for our scheme as shown in Fig \ref{f:sensing} where the sensitivity with positive detuning is worse than that with negative detuning. This may come from the shape of the ODMR without the RF. We cannot perfectly fit the shape by a sum of Lorentzians due to the assymetric shape where gradient of the contrast of the positive detuning is different from that of the negative detuning, as shown in the Fig. \ref{f:ODMRwith/outRF}(a). Importantly, such an assymetric shape of the ODMR without applying DC magnetic fields was discussed in please cite my JPCM paper in 2016.\cite{matsuzaki2016optically} However, in our case, we apply orthogonal DC magnetic fields, which is slightly different from the setup in please cite my JPCM paper in 2016.\cite{matsuzaki2016optically} We leave the detailed study of these for a future work.

Finally, we discuss the frequency tunability of our sensing scheme. We tune the detectable frequency of the target field by changing the $\mathrm{RF_{control}}$ amplitude (i.e., $\omega_{\mathrm{RFt}}=2E'_x \pm \gamma_e B_{\mathrm{RFc}}$). However, a strong $\mathrm{RF_{control}}$ amplitude could lead to the violation of the RWA, as we mentioned in the previous section. The RWA will be invalid when the driving amplitude becomes comparable to half of the resonant frequency. Thus, it would be practical to detect the target field between 4.24~MHz ($\simeq E'_x/2$, lower limit) and 12.7~MHz ($\simeq 3E'_x/2$, upper limit).\cite{scheuer2014precise}

\section{CONCLUSION}
In conclusion, we propose and demonstrate frequency-tunable magnetic field sensing based on CW-ODMR using RF double-dressed states of NV centers in diamond. We can tune the detectable frequency of the  magnetic fields by adding control RF fields, which is stark contrast to the previous method where the detectable frequency is fixed. Using our setup, the sensitivity achieved by our scheme is higher than that of the previous scheme for target AC magnetic fields with frequencies range under 7~MHz or above 10~MHz. We also found that a hybrid scheme combining the previous scheme and our scheme can improve the bandwidth by approximately 3.70~MHz. These results contribute to the development of practical AC magnetic field sensors.

Finally, we discuss possible future work.
To broaden the frequency range, we can apply DC electric fields that plays a role to increase the energy splitting between \(|B\rangle\) and \(|D\rangle\) \cite{Dolde2011}. In this case, we can increase $B_{\rm{RF}_{\rm{C}}}$ amplitude without violating the RWA, and we should be able to detect AC magnetic field with larger or smaller frequencies.
Alternatively we can increase the orthogonal DC magnetic fields, which also increases the energy splitting between the first excited state and second exited state of the Hamiltonian.
However, in this case, the eigenstates of the Hamiltonian are not described by \(|B\rangle\) and \(|D\rangle\), and so further analysis is required to check the validity of our method.

\begin{acknowledgments}
This work was supported by MEXT Q-LEAP (Grant No. 337 JPMXS0118067395), MEXT KAKENHI (Grant Nos. 19H05826, 20H05661, 22H01558 and 22K03524), and CSRN, Keio University. This work was also supported by Leading Initiative for Excellent Young Researchers MEXT Japan and JST presto (Grant No. JPMJPR1919) Japan, JST (Moonshot R\&D)(Grant Number JPMJMS226C), and Kanazawa University CHOZEN Project 2022.
\end{acknowledgments}

\section*{AUTHOR DECwLARATIONS}
\subsection*{Conflict of Interest}
The authors have no conflicts to disclose.
\subsection*{Author Contributions}

\section*{DATA AVAILABILITY}
The data that support the findings of this study are available from the corresponding authors upon reasonable request.

\appendix
\section{Details of Our Previous Sensing Method}
\label{appendix}
In this section, we provide details about the previous sensing method.\cite{Saijo2018,Yamaguchi2019} In this method, only one RF field which is used to generate the RF dressed states, and this field is also the target AC magnetic field. Based on Eqs.~(\ref{eq:HNV})-(\ref{eq:E'x}), the total Hamiltonian ($\hat{H}^{\mathrm{p}}_{\mathrm{tot}}$) of the NV centers with the RF field and MW is given as follows:
\begin{align*}
\hat{H}^{\mathrm{p}}_{\mathrm{tot}} \simeq &  D'\hat{S}_z^2 + E_x'(\hat{S}_x^2 -\hat{S}_y^2)\\
&+{\gamma}_eB_{\mathrm{RF}} \hat{S}_z\cos (\omega_{\mathrm{RF}}t)\\
&+{\gamma}_eB_{\mathrm{MW}} \hat{S}_x \cos (\omega_{\mathrm{MW}} t),
\stepcounter{equation}\tag{\theequation} 
\end{align*}
where, $B_{\mathrm{RF}}$ is amplitude of the target RF field along the $z$ direction and $\omega_{\mathrm{RF}}$ is the frequency of the target RF field. By moving to a rotating frame $\hat{U}^{\mathrm{p}} = e^{i(\omega_{\mathrm{RF}}|D\rangle \langle D|-\omega_{\mathrm{MW}}|0\rangle \langle 0|)t}$, the effective Hamiltonian ($\hat{H}^{\mathrm{p}}_{\mathrm{eff}}$) 
after using the RWA for the MW and RF field is given as
\begin{align*}
\hat{H}^{\mathrm{p}}_{\mathrm{eff}} \simeq &  (D'-\omega_{\mathrm{MW}}+E'_x)|B\rangle \langle B|\\
&+(D'-\omega_{\mathrm{MW}}-E'_x+\omega_{\mathrm{RF}})|D\rangle \langle D|\\
&+\frac{1}{2}\gamma_eB_{\mathrm{MW}} (|B\rangle \langle 0|+|0\rangle \langle B|)\\
&+\frac{1}{2}\gamma_eB_{\mathrm{RF}} (|B\rangle \langle D|+|D\rangle \langle B|).
\stepcounter{equation}\tag{\theequation} 
\label{eq:doubleRWA}
\end{align*}
As already noted in the main text, We treat this system as a coupled harmonic oscillator and we can rewrite Eq.~(\ref{eq:doubleRWA}) as
\begin{align*}
\hat{H}^{\mathrm{p}}_{{\mathrm{eff}}} \simeq &\omega ^{\mathrm{p}}_ { b } \hat { b } ^ { \dagger } \hat { b } 
+\omega ^{\mathrm{p}}_ { d } \hat { d } ^ { \dagger } \hat { d }\\
&+\lambda^{\mathrm{p}} \left(\hat { b } ^ { \dagger } + \hat { b } \right)
+J^{\mathrm{p}}\left(\hat { b } \hat { d } ^ { \dagger }+\hat { b } ^ { \dagger } \hat { d } \right),
\stepcounter{equation}\tag{\theequation} 
\end{align*}
where $\omega ^{\mathrm{p}}_b = D'-\omega_{\mathrm{MW}}+E'_x$, $\omega ^{\mathrm{p}}_d = D'-\omega_{\mathrm{MW}}-E'_x+\omega_{\mathrm{RF}}$, $\lambda^{\mathrm{p}} = \frac{1}{2}\gamma_eB_{\mathrm{MW}}$ and  $J^{\mathrm{p}} = \frac{1}{2}\gamma_eB_{\mathrm{RF}}$. In addition, we define \(\hat{b}^{\dagger} = |B\rangle \langle 0|\)  \((\hat{d}^{\dagger}=|D\rangle \langle 0|)\) as a creation operator and \(\hat{b} = |0\rangle \langle B|\)  \((\hat{d}=|0\rangle \langle D|)\) as an annihilation operator.

From the Heisenberg equation, we can obtain the probability of the state in $|0\rangle$ given by in the same form as Eqs.~(\ref{eq:p0})-(\ref{eq:d}). Finally, we can calculate the resonant MW frequencies ($\omega^{\mathrm{p}}_{\mathrm{res1\pm}}$) as follows. 
\begin{align}
\omega^{\mathrm{p}}_{\mathrm{res1\pm}} = D' + \frac{1}{2}\omega_{\mathrm{RF}}\pm{\sqrt{(2E'_x-\omega_{\mathrm{RF}})^2+(\gamma_e B_{\mathrm{RF}})^2}}.
\label{eq:doublereso1}
\end{align}

Moreover, note that we calculate the similar process and obtain another resonant MW frequency when considering MW applied along the $y$ direction, as following formula. On the other hand, if we apply the MW along $y$ direction, we obtain
\begin{align}
\omega^{\mathrm{p}}_{\mathrm{res2\pm}} = D' - \frac{1}{2}\omega_{\mathrm{RF}}\pm{\sqrt{(2E'_x-\omega_{\mathrm{RF}})^2+(\gamma_e B_{\mathrm{RF}})^2}},
\label{eq:doublereso2}
\end{align}
as the resonant frequencies.

These results indicate that the resonant frequency 
has a linear dependence on the amplitude of
changes linearly according to the amplitude value of
the target RF field when
$\omega_{\mathrm{RF}}=2E'_x$ is satisfied. Based on the above theory, we obtain the contrast change ($S$) derived from the target RF field amplitude while fixing the MW frequency at $D' + \frac{1}{2}\omega_{\mathrm{RF}}$ or $D' - \frac{1}{2}\omega_{\mathrm{RF}}$ 
in actual experiment. Therefore, we can measure the 
AC magnetic field from the change in the contrast.

\nocite{*}
\section*{REFERENCES}
\bibliography{aipsamp}

\end{document}